\documentclass{aastex}

\usepackage{amsmath} 
\usepackage{natbib}
\usepackage{pdflscape}
\usepackage{natbib}
\setcitestyle{round,aysep={},yysep={;}}

\usepackage{lineno}
\slugcomment{}

\newcommand{\be}{\begin{equation}}
\newcommand{\ee}{\end{equation}}

\shorttitle{Exact rotation curves with MOND}
\shortauthors{L\'opez-Corredoira \& Betancort-Rijo}

\begin{document}

\title{Exact semianalytical calculation of rotation curves with Bekenstein--Milgrom nonrelativistic MOND}
\author{M.  L\'opez-Corredoira\altaffilmark{1,2}, J. E. Betancort-Rijo\altaffilmark{1,2}}
\altaffiltext{1}{Instituto de Astrofisica de Canarias, E-38205 La Laguna, Tenerife, Spain; fuego.templado@gmail.com}
\altaffiltext{2}{Departamento de Astrofisica, Universidad de La Laguna, E-38206 La Laguna, Tenerife, Spain}

\begin{abstract}
Astronomers use to derive MOdified Newtonian Dynamics (MOND) rotation curves using the simple
algebraic rule of calculating the acceleration as equal to the Newtonian acceleration ($a$) divided by some 
factor $\mu (a)$.
However, there are velocity differences between this simple rule and the calculation
derived from more sophisticated MOND versions such as AQUAL or QMOND, created to 
expand MOND heuristic law and preserve the conservation of momentum,  angular momentum, and energy, and follow the weak equivalence principle.
Here we provide recipes based on Milgrom's proposal to calculate semianalytically (without numerical simulations)
MOND rotation curves for any density distribution  based on AQUAL, applying it to different models of thin disks.
The application of this formalism is equivalent to the creation of a fictitious phantom mass whose field may be used in a Newtonian way to calculate iteratively the MOND accelerations.
In most cases, the differences between the application of the simple algebraic rule and the AQUAL-MOND calculations are small, $\lesssim 5$\%. However, the error of the algebraic solution is larger than 5\% when more than half of the mass is in the MONDian regime (where Newtonian and MOND rotation speeds differ by more than 10\%), reaching in some cases $>70$\% discrepance, such as in Maclaurin disks, representative of galaxies for which the rotational velocity rises to the edge of the disk as is seen in irregular galaxies.
The slope of the rotation speed in the dependence with the radius or the vertical distance of the plane is also significantly changed.
%The concept of phantom mass can be extended to cosmology and large-scale structure, which is not
%the main aim of this paper. At present, with simple order of magnitudes calculations, we derive that formation of aggregates of gravitational masses of $\sim 3\times 10^{15}$ M$_\odot $, typical of rich cluster of galaxies, are enough to make $\Omega =1$ in a MOND Universe without dark matter and dark energy.
%Therefore, the calculation of MOND with the exact solution is important when using high precision data,
%especially to examine small subtleties that might allow the rejection of MOND or dark matter hypotheses. Astronomers should evaluate the ratio of disk mass within MONDian regime and, in cases of being $>0.5$, use the exact solution for an accuracy better than 5\%.
\end{abstract}

%\keywords{galaxies: kinematics and dynamics -- dark matter -- galaxies: spiral}
{\it Unified Astronomy Thesaurus concepts:} Spiral galaxies (1560); Galaxy dynamics (591).

%________________________________________________________________

\section{Introduction}
\label{.intro}

One of the most challenging alternatives to dark matter hypotheses to explain rotation
curves in spiral galaxies and other astrophysical observations
is the modification of gravity laws proposed in Modified Newtonian Dynamics (MOND; 
\citet{Mil83a,Fam12,San15}), which modifies the Newtonian laws for accelerations lower than 
$a_0\sim 1\times 10^{-10}$ m/s$^2$. 
The value of the acceleration scale, $a_0$, defining the variation with respect to Newton's law necessary to fit the rotation curves, is very similar in all of the galaxies, and it has been interpreted as a possible sign of confirmation of MOND \citep{Lel17}. \citet{Bot15}, \citet{Rod18} and \citet{Zob20} find
some small variations of $a_0$ in different galaxies invalidating the universal application of a MOND-like modification of gravity with constant $a_0$, but this result was found to 
rely on galaxies with very uncertain distances and/or nearly edge-on orientations \citep{Kro18}.

The hypothesis of MOND has been used in many contexts of astrophysics in the past almost 40 yr \citep{Fam12}, becoming stronger in many aspects and weaker in other aspects. Without further assumptions, it cannot compete with $\Lambda $CDM to explain the large-scale structure and other cosmological predictions \citep{Fam12}. However, within the galactic scales, it is becoming a strong competitor, particularly, as mentioned,
in the explanation of the rotation curves of spiral galaxies. There are some
features of the rotation curves that are not well explained by MOND, for instance, the variation
of the amplitude of the rotation speed with the height from the plane \citep{Lis19}, but 
in the plane it works pretty well \citep[e.g.,][]{Beg91,San96,Bot15}.

The theory of MOND violates the strong equivalence principle, but not necessarily the weak equivalence principle \citep{Mil83a,Mil83b,Mil83c}. MOND was in principle a phenomenological approach \citep{Mil83a,Mil83b,Mil83c}, but some elements  were also incorporated that make it compatible with more general gravitation theories. The AQUAdratic Lagrangian theory (AQUAL; \citet{Bek84}) expanded MOND to preserve the conservation of momentum, angular momentum, and energy, and follow the weak equivalence principle. Another latter proposal
is quasi-linear formulation of MOND (QUMOND; \citet{Mil10}), which provides different solutions
to AQUAL in asymmetric systems, although the two-body force in the deep-MOND limit is the same
\citep{Zha10}. Also, a relativistic gravitation theory of MOND would be developed under the name Tensor-Vector-Scalar (TeVeS; \citep{Bek04}), which also tried to provide consistency with certain cosmological observations, including gravitational lensing.

Modification of the dynamics in order to reproduce Milgrom's heuristic law while still benefiting from usual conservation laws such as the conservation of momentum starts from the action at the classical level: either by a modification of the Newtonian second law (of inertia), 
$\mathbf{F}=m\mu(a)\mathbf{a}$, or by modifying the Newtonian gravitation attraction $\mathbf{F}=\frac{G\,M\,m\mathbf{r}}{\mu (a)r^3}$ \citep[Sect. 6]{Fam12}.
The modification of the law of inertia also implies modification of the laws in the case of electromagnetic forces. A first attempt of a test to differentiate between both scenarios with data of galaxies was recently 
carried out by \citet{Pet20}, with results favoring the modified gravity interpretation.

This MOND modelling, which is more sophisticated than the algebraic expression, is used in some cases to study the dynamics and evolution of disk galaxies  \citep[e.g.,][]{Tir07,Tir08}, but most
astronomers \citep[e.g.,][]{Beg91,San96,Bot15,Lis19} are still using the simple
algebraic rule of calculating the MOND force as equal to the Newtonian one divided by some factor $\mu $.
It is known (\citet{Bra95}, \citet[\S 6.5.1]{Fam12}) that there are small 
velocity differences between the simple rule and the exact\footnote{Assuming AQUAL or QMOND as exact.
With ``exact,'' we mean here the solution of their mathematical equations without any approximation. 
Whether they represent the exact description of the modified gravity or not is another question.} 
calculation;
\citet[Equation (25)]{Bra95} also proposed an approximate analytical expression, which was
used, for instance, by \citet{Pet20} in their analyses of rotation curves, although still with some
significant differences with respect to the exact solution.
It has been shown that the maximum difference between formulations is on the order of 10\%  in the case of an exponential disk or Kuzmin disk, and similarly with QUMOND \citep{Ban18b}. However, other types  of density distributions, with larger amounts of mass in the outer parts of the disk, have not been explored so far and may have larger differences, and there is no useful algorithm that can be used by astronomers for the application of the exact solution.
$N$-body and hydrodynamic codes that solve the modified Poisson equation either of AQUAL or QUMOND were also developed
by several authors \citep{Bra99,Tir07,Lon09,Ang12,Can15,Lug15,Ban18b}; however no recipe in analytical terms was given to reach these exact MOND solutions.

Here we calculate ``exact MOND'' rotation curves for any density distribution 
(\S \ref{.exact})  based on AQUAL \citep{Bek84} formulation of MOND, and we will carry out an extended analysis of these exact calculations for different models of disks (\S \ref{.application}).
We will provide recipes as originally proposed by \citet{Mil86}), in a semianalytical method: without Poisson solver codes through numerical simulations, but with an iterative method of integrals' calculations that
are solved numerically. In \S \ref{.phanton}, 
we analyze the interpretation of MOND gravity in terms of an equivalent phantom 
density field and discuss the consequences this may bring to MOND interpretations and the different controversies about its applicability that can be solved by a correct application of the theory.
Conclusions are summarized in \S \ref{.conclus}.

\section{Exact calculation of MOND rotation curve}
\label{.exact}

\subsection{Newtonian gravity}
\label{.Newton}

In Newtonian gravity, a density distribution $\rho (\mathbf{r})$ produces a field of accelerations
\begin{equation}
\label{acc_newton}
\mathbf{g}_{\rm N}[\rho ](\mathbf{r})=G\int d\mathbf{r'} \frac{\rho (\mathbf{r'})}
{|\mathbf{r'}-\mathbf{r}|^3} (\mathbf{r'}-\mathbf{r})
.\end{equation}

Using cylindrical coordinates
in which $R$ and $z$ are radial and vertical distance, respectively, and $\phi $ is the azimuthal angle,
in an axisymmetric matter distribution ($\rho \ne \rho (\phi )$),
the rotation speed $V_c$ for stars in equilibrium with the centrifugal force with Newtonian gravitation is 
 \citep[Appendix A]{Chr20}

\begin{equation}
g_{{\rm N,R}}(R,z)=\frac{-V_c^2}{R}=
\frac{-2\,G}{R} \int _{-\infty}^\infty dz'\int_0^\infty dR'\,R'\rho (R',z')
\end{equation}\[
\ \ \ \ \ \ \ \ \ \ \ \ \times [C_E\,E(k)+C_K\,K(k)]
\]\[
C_E=\frac{(R'+R)(R'-R)+(z-z')^2}{[(R'-R)^2+(z-z')^2]\sqrt{(R'+R)^2+(z-z')^2}}
\]\[
C_K=-\frac{1}
{\sqrt{(R'+R)^2+(z-z')^2}}
\]\[
k=\sqrt{\frac{4\,R\,R'}{(R'+R)^2+(z-z')^2}}
,\]
where  $K(k)$ and $E(k)$ are the complete elliptical integrals of the first and second kind, respectively, and
 $g_{{\rm N,R}}$ is the Newtonian radial acceleration. 

The vertical acceleration with Newtonian gravity is
\begin{equation}
g_{{\rm N,z}}(R,z)=\frac{-G\,z}{2\,R^{3/2}}\int _{-\infty}^\infty
dz'\int _0^\infty dR'\frac{\rho (R',z')}{\sqrt{R'}}H(k)\end{equation}
\[H(k)=\int_0^{2\pi} d\phi '\frac{1}{[(2/k^2)-\cos{\phi '}]^{3/2}}
\]\[
k=\sqrt{\frac{4\,R\,R'}{(R'+R)^2+(z-z')^2}}
.\]

\subsection{MOND gravity, with the algebraic rule}

The general algebraic rule of Milgrom's empirical law
usually applied by astronomers to calculate rotation curves with MOND is as follows
\citep[Equation 7]{Fam12}
\begin{equation}
\mathbf{g}_{\rm M-R}=\frac{\mathbf{g}_{\rm N}}{\mu (|\mathbf{g}_{\rm M-R}|/a_0)}
\label{MONDrule}
,\end{equation}
where $\mu (x)$ is an interpolating function. The standard interpolating function is
\begin{equation}
\label{mu}
\mu (x)=\sqrt{\frac{1}{1+x^{-2}}}
.\end{equation}
With the last two equations,
\begin{equation}
|g_{\rm M-R}|=\sqrt{\frac{1}{2}g_{\rm N}^2+\sqrt{\frac{1}{4}g_{\rm N}^4+
g_{\rm N}^2a_0^2 }}
,\end{equation}\[
\mathbf{g}_{\rm M-R}=\mathbf{g}_{\rm N}\frac{|g_{\rm M-R}|}{|g_{\rm N}|}
,\]
as used, for instance, by \citet{Bot15} and \citet{Lis19} for rotation curve fits. The rotation speed in the plane would be $V_c=\sqrt{|\mathbf{g}_{\rm M-R}|\,R}$.

\subsection{Exact MOND gravity}

Previous Eq. (\ref{MONDrule}) leads to some consistency problems. For instance, in a two-body case, as the implied force is not symmetric in the two masses, Newton’s third law does not hold, so the momentum is not conserved \citep{Mil83a,Fel84,Fam12}. Precisely because of this,
more sophisticated formulations of MOND were created as a modification of classical dynamics: the AQUAL
model by \citet{Bek84} that we will use here.

An unpleasant characteristic of MOND is the nonlinear nature of its equations. However, nonlinear field equations can be
expressed as linear field equations containing a self-source term. This may be convenient for heuristic reasons
and because in this manner the equations can be solved by well-known linear methods. Obviously, this does not
mean that the nonlinear difficulties can be avoided, because, due to the presence of the self-source term, the equations
must be solved self-consistently. However, as we shall show, this solution may be obtained with considerable precision in
a few iterations.

The Poisson equation of the AQUAL solution for the potential $\phi $ is
\begin{equation}
\nabla (\mu (x)\nabla \phi )=4\pi G \rho 
,\end{equation}
\[x=\frac{|\mathbf{g}_{\rm M-E}|(\mathbf{r})}{a_0}
,\]
where $\mathbf{g}_{\rm M-E}(\mathbf{r})$ is the exact MOND acceleration.
The left-hand side of this equation may be written in the form $\mu (x) \nabla ^2\phi +(\nabla \mu(x))(\nabla \phi)$. Since 
\begin{equation}
\nabla \mu (x)=\frac{1}{a_0}\frac{d\mu }{dx}(x).\nabla |\mathbf{g}_{\rm M-E}|
\end{equation}
and
\begin{equation}
\mathbf{g}_{\rm M-E}=-\nabla \phi
,\end{equation}
we get
\begin{equation}
\mu (x)\nabla ^2 \phi =4\pi G \rho + \frac{1}{a_0}\frac{d\mu }{dx}(x)\nabla |\mathbf{g}_{\rm M-E}|\,\cdotp\mathbf{g}_{\rm M-E}
.\end{equation}
The second term in this equation is the self-source term, which can be interpreted as a density of phantom matter.
Therefore, the solution is
\begin{equation}
\phi (\mathbf{r})=\int d\mathbf{r'} \frac{G\rho (\mathbf{r'})+\frac{1}{4\pi a_0}\frac{d\mu }{dx}(\mathbf{r'})
\nabla |\mathbf{g}_{\rm M-E}(\mathbf{r'})|\,\cdotp \mathbf{g}_{\rm M-E}(\mathbf{r'})}
{\mu (\mathbf{r'})|\mathbf{r}-\mathbf{r'}|}
.\end{equation}
For a thin disk, the exact calculation of MOND acceleration  $\mathbf{g}_{\rm M-E}(\mathbf{r})$
is equivalent to the Newtonian gravity calculation (equations given in \S \ref{.Newton}), but setting as density
\begin{equation}
\label{rhoM-E}
\rho ^*(\mathbf{r})=\frac{\rho (\mathbf{r})}{\mu (x)}+\frac{1}{4\pi\,G\,a_0\,\mu(x)}
\frac{d\mu }{dx}(x)
\end{equation}\[
\times \left|g_{\rm M-E,R}(\mathbf{r})\frac{\partial g_{\rm M-E}(\mathbf{r})}{\partial R}
+g_{\rm M-E,z}(\mathbf{r})\frac{\partial g_{\rm M-E}(\mathbf{r})}{\partial z}\right|
.\]
This equation was also derived by \citet[Equation 4]{Mil86}.
That is, we can calculate the acceleration iterating between Eq. (\ref{acc_newton}) $\mathbf{g}_{\rm M-E}[\rho ](\mathbf{r})=\mathbf{g}_{\rm N}[\rho ^*](\mathbf{r})$ and Eq. (\ref{rhoM-E}). 
For the first iteration, we set $\mathbf{g}_{\rm M-E}=\mathbf{g}_{\rm M-R}$.
In practice, when $g_{\rm M-E}$ is close to $g_{\rm M-R}$, we only need two iterations (considering the
first iteration $\rho ^*=\rho $), since
$\mathbf{g}_{\rm M-E}[\rho ](\mathbf{r})\approx \mathbf{g}_{\rm N}[\rho ^*[\mathbf{g}_{\rm M-R}]](\mathbf{r})$ is a good approximation. We will show in \S \ref{.maclaurin} that even among cases with very large
differences between $g_{\rm M-E}$ and $g_{\rm M-R}$, the iterative process converges quickly in 
three to four iterations.

The rotation speed in the plane would again be $V_c=\sqrt{|\mathbf{g}_{\rm M-E}|\,R}$.

\section{Application to different thin-disk density models}
\label{.application}

In a thin disk, 
$\rho $ is zero in off-XY-plane regions and there is a surface density
$\sigma (R)$ such that
\begin{equation}
\sigma (R)\equiv \int _{-\infty}^\infty dz\,\rho(R,z)
.\end{equation}
In this thin disk, we avoid the integration in the vertical direction
by setting $z'=0$ in the previous equations.
The mass within radius $R$ of this thin disk is
\begin{equation}
M(R)=2\pi \int _0^RdR'R'\sigma (R')
\end{equation}

Considerations of a thick disk may be found in the literature too \citep[e.g.,][]{Cas83}.

\subsection{Central point-like mass}
\label{.central}

\begin{figure}
\vspace{0cm}
\centering
\includegraphics[width=9cm]{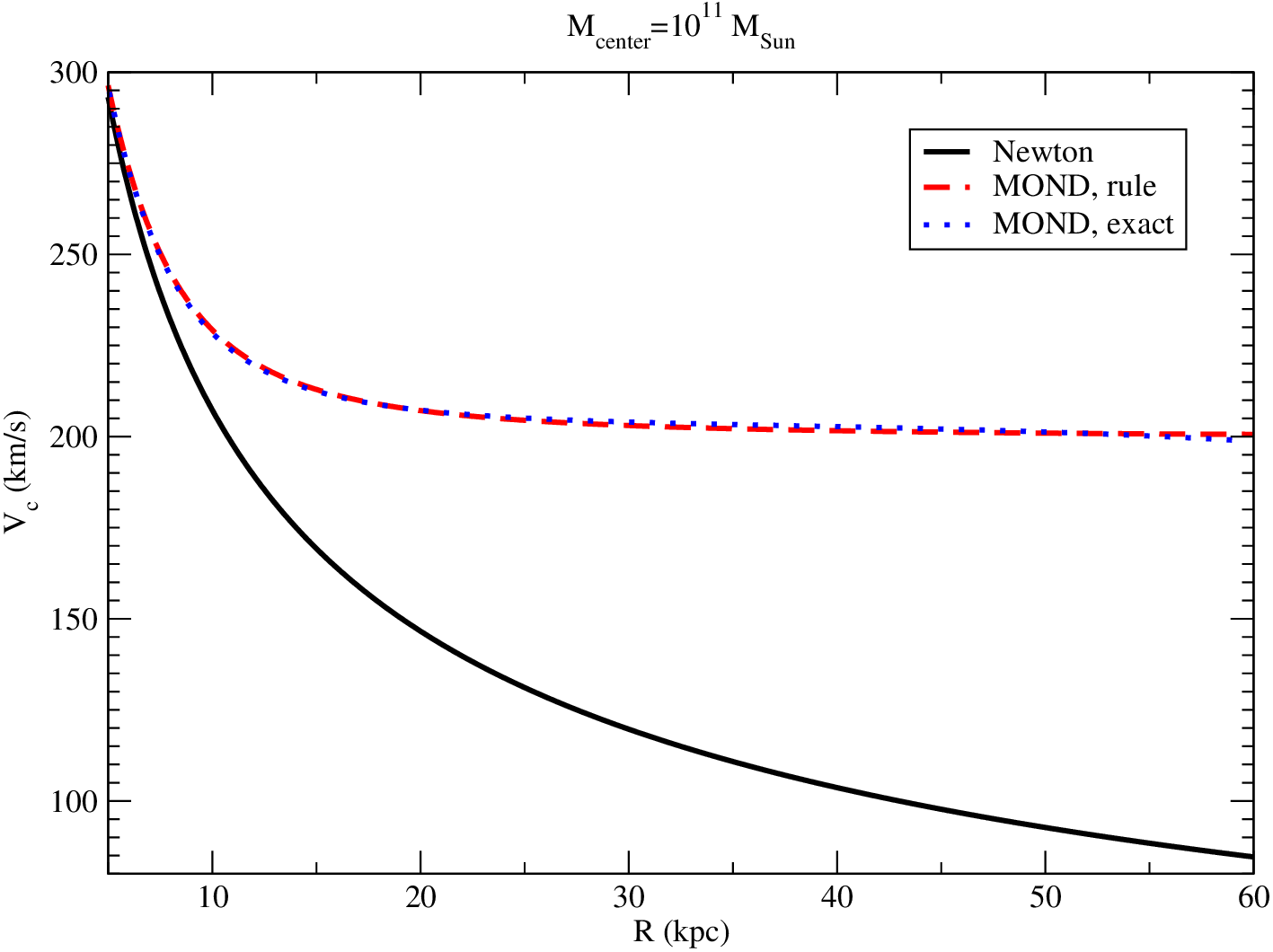}
\caption{Rotation curve in the plane ($V_c=\sqrt{|g_{\rm R}|\,R}$ at $z=0$) for central point-like mass of $M=10^{11}$ $M_\odot $ calculated with Newtonian gravity ($\mathbf{g}_{\rm N}$), MOND-rule ($\mathbf{g}_{\rm M-R}$), and MOND-exact ($\mathbf{g}_{\rm M-E}$), respectively.}
\label{Fig:MONDexact1}
\end{figure}

By construction, $\mathbf{g}_{\rm M-E}[\rho ](\mathbf{r})=\mathbf{g}_{\rm M-R}[\rho ](\mathbf{r})$ in cases
of spherical symmetry in the density distribution.
We will test it in the simplest case, a point-like mass in the center of the galaxy:

\begin{equation}
\sigma (R)=\left \{ 
\begin{array}{ll}
M\delta (R),& \mbox{$R=0$} \\
0,& \mbox{$R>0$} 
\end{array}
\right \} \;.
\end{equation}
In Fig. \ref{Fig:MONDexact1}, we show this perfect agreement (the very slight relative differences lower than 1\% are due to numerical calculation errors) of both MOND algorithms in counter-distinction of Newtonian gravity.

\subsection{Exponential disk}

\begin{figure}
\vspace{0cm}
\centering
\includegraphics[width=9cm]{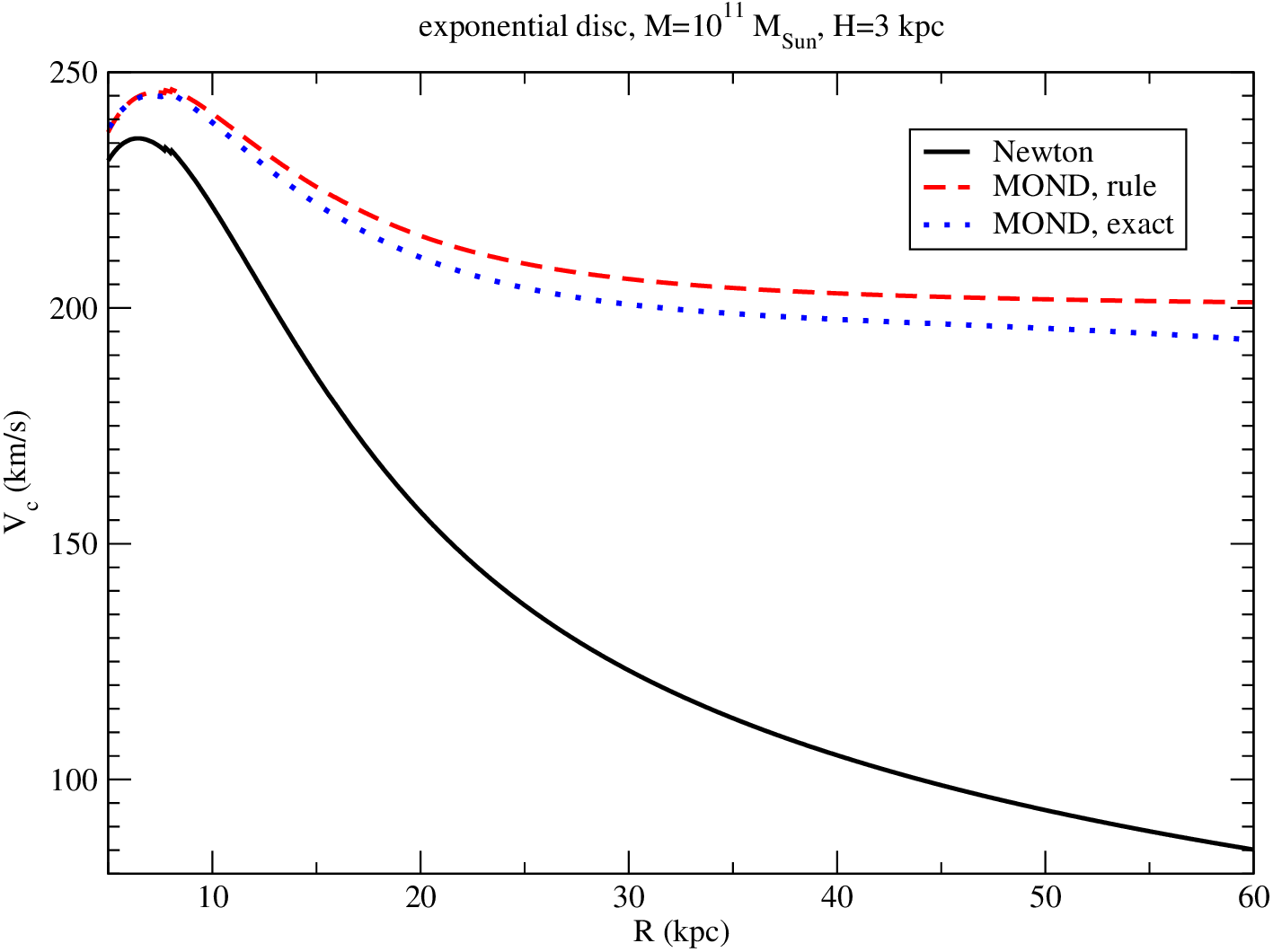}
\caption{Rotation curve in the plane ($V_c=\sqrt{|g_{\rm R}|\,R}$ at $z=0$)  for exponential disk  of $M=10^{11}$ $M_\odot $, $H=3$ kpc, calculated with Newtonian gravity ($\mathbf{g}_{\rm N}$), MOND-rule ($\mathbf{g}_{\rm M-R}$), and MOND-exact ($\mathbf{g}_{\rm M-E}$), respectively.}
\label{Fig:MONDexact2}
\end{figure}

The most usual fit of the galactic thin disk is with a simple exponential law:

\begin{equation}
\sigma (R)=\frac{M}{2\pi H^2}\exp{\left(-\frac{R}{H}\right)}
,\end{equation}
where $M$ is the total mass of the disk, and $H$ is its scale length.
In Figs. \ref{Fig:MONDexact2} and \ref{Fig:MONDexact2}, we show the rotation curves for a total mass
of $10^{11}$ $M_\odot $, scale lengths of 3 and 6 kpc, for Newtonian gravity ($\mathbf{g}_{\rm N}$), MOND-rule ($\mathbf{g}_{\rm M-R}$) and MOND-exact ($\mathbf{g}_{\rm M-E}$), respectively.

\begin{figure}
\vspace{0cm}
\centering
\includegraphics[width=9cm]{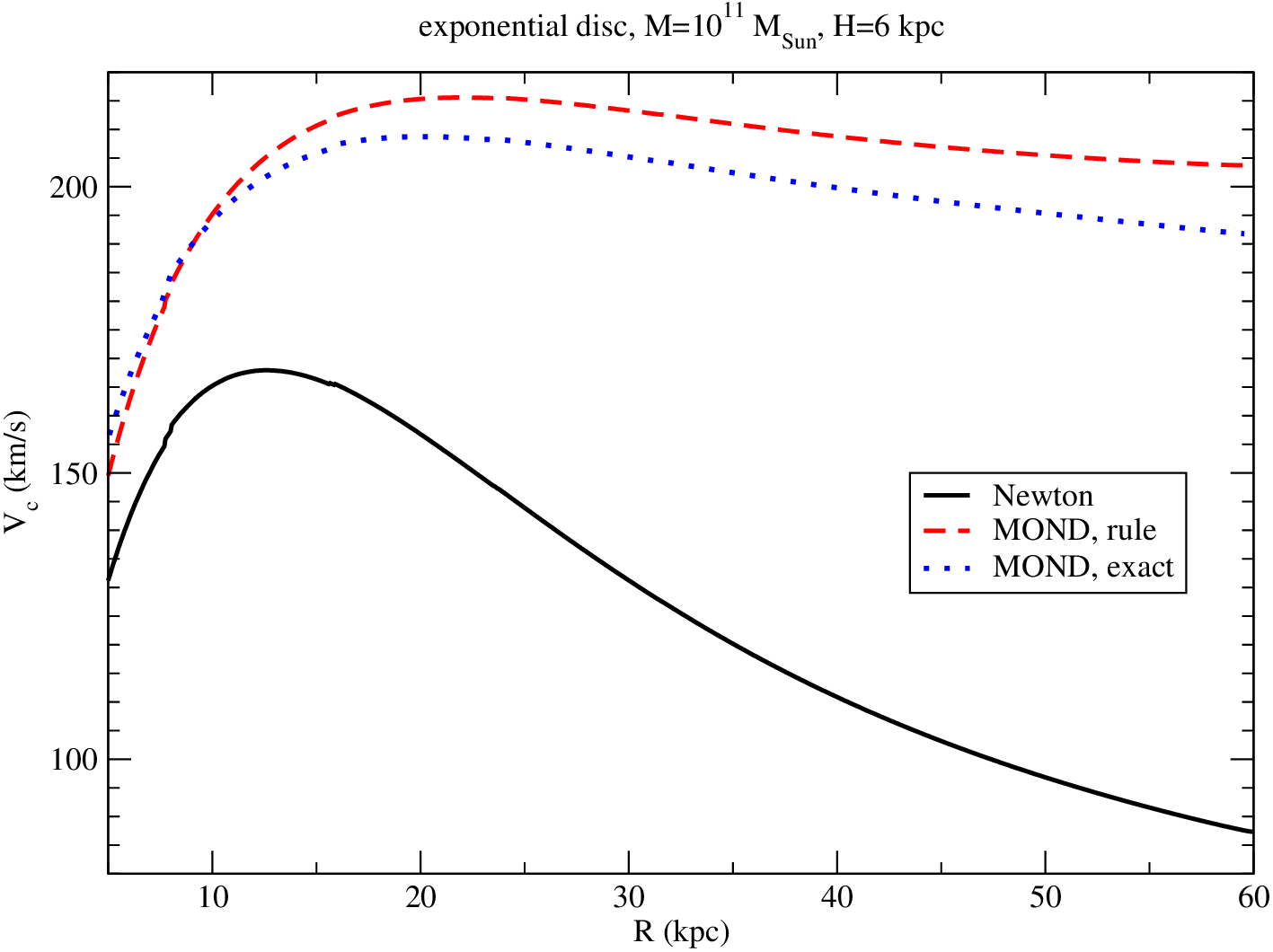}
\caption{Rotation curve in the plane ($V_c=\sqrt{|g_{\rm R}|\,R}$ at $z=0$) for exponential disk of $M=10^{11}$ $M_\odot $, $H=6$ kpc, calculated with Newtonian gravity ($\mathbf{g}_{\rm N}$), MOND-rule ($\mathbf{g}_{\rm M-R}$), and MOND-exact ($\mathbf{g}_{\rm M-E}$), respectively.}
\label{Fig:MONDexact3}
\end{figure}

\begin{figure}
\vspace{0cm}
\centering
\includegraphics[width=9cm]{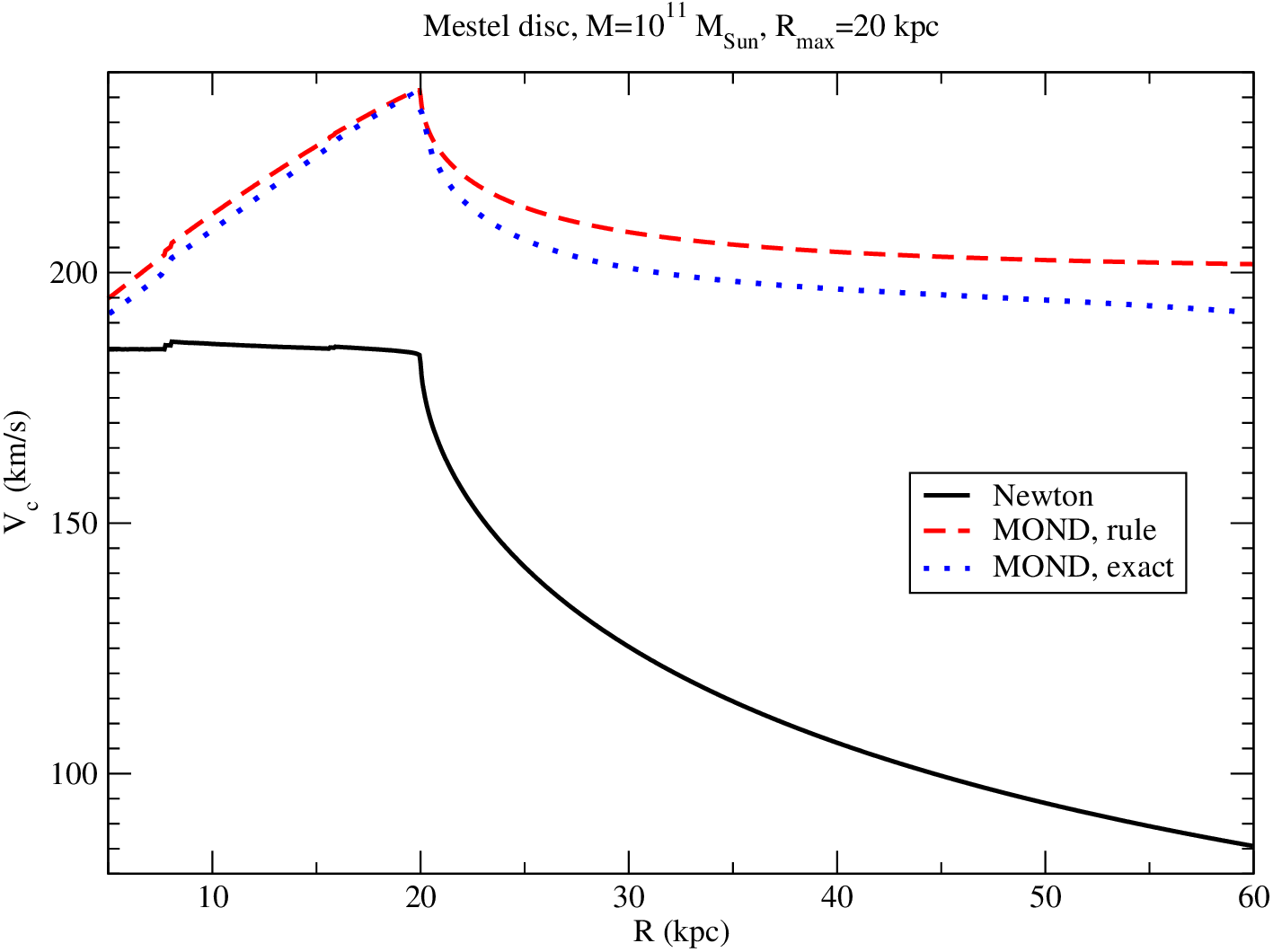}
\caption{Rotation curve in the plane ($V_c=\sqrt{|g_{\rm R}|\,R}$ at $z=0$) for a Mestel disk of $M=10^{11}$ $M_\odot $, $R_{\rm max}=20$ kpc, calculated with Newtonian gravity ($\mathbf{g}_{\rm N}$), MOND-rule ($\mathbf{g}_{\rm M-R}$), and MOND-exact ($\mathbf{g}_{\rm M-E}$), respectively.}
\label{Fig:MONDexact6}
\end{figure}

\subsection{Mestel disk}

A distribution characterized by giving a flat rotation curve in Newtonian gravity without any
extra component, such as dark matter halo, is the Mestel disk, whose dependence with the
galactocentric radius goes as \citep{Sch12}
\begin{equation}
\sigma (R)=\left \{ 
\begin{array}{ll}
\frac{M}{2\pi R_{\rm max}R}{\rm arccos}
\left(\frac{R}{R_{\rm max}}\right),& \mbox{$R\le R_{\rm max}$} \\
0,& \mbox{$R>R_{\rm max}$} 
\end{array}
\right \} \;.
\end{equation}
$R_{\rm max}$ is the radius of the disk. For larger radii, there is not any mass.
In Fig. \ref{Fig:MONDexact6}, we show the rotation curve for a total mass
of $10^{11}$ $M_\odot $ and maximum radius of 20 kpc, for Newtonian gravity ($\mathbf{g}_{\rm N}$), MOND-rule ($\mathbf{g}_{\rm M-R}$) and MOND-exact ($\mathbf{g}_{\rm M-E}$) respectively.

\subsection{Maclaurin disk}
\label{.maclaurin}

\begin{figure}
\vspace{0cm}
\centering
\includegraphics[width=9cm]{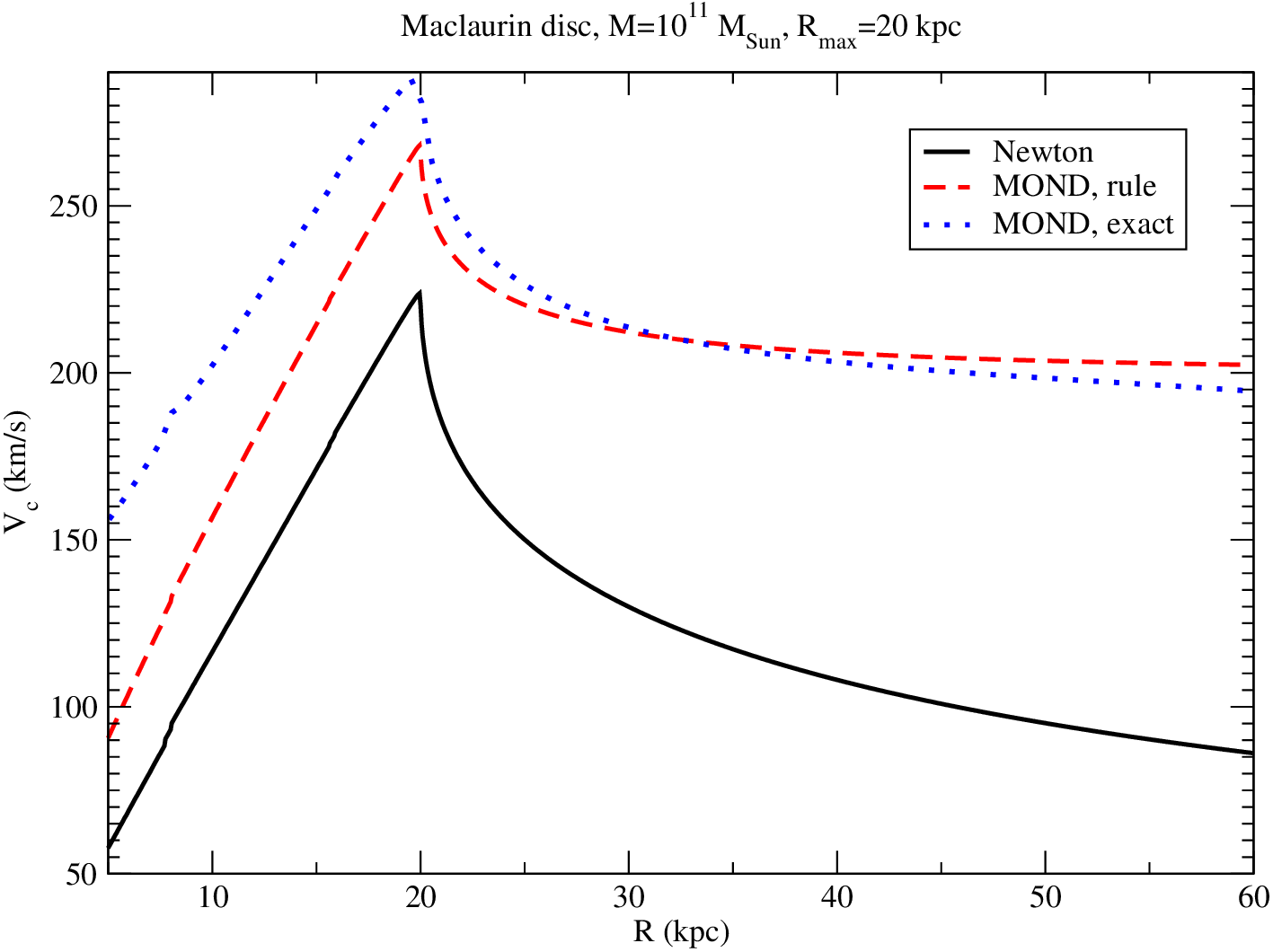}
\caption{Rotation curve in the plane ($V_c=\sqrt{|g_{\rm R}|\,R}$ at $z=0$) for a Maclaurin disk of $M=10^{11}$ $M_\odot $, $R_{\rm max}=20$ kpc, calculated with Newtonian gravity ($\mathbf{g}_{\rm N}$), MOND-rule ($\mathbf{g}_{\rm M-R}$), and MOND-exact ($\mathbf{g}_{\rm M-E}$) (only with two iterations of the Equations (\ref{acc_newton}) and
(\ref{rhoM-E})), respectively.}
\label{Fig:MONDexact7}
\end{figure}

The Maclaurin disk, a limiting case of the Maclaurin spheroid, is applicable to the study of spiral galaxies for which the rotational velocity rises to the edge of the disk as is seen in irregular galaxies \citep{Sch09}.
It follows
\begin{equation}
\sigma (R)=\left \{ 
\begin{array}{ll}
\frac{3\,M}{2\pi\,R_{\rm max}^2}\sqrt{1-\left(\frac{R}{R_{\rm max}}\right)},& \mbox{$R\le R_{\rm max}$} \\
0,& \mbox{$R>R_{\rm max}$} 
\end{array}
\right \} \;.
\end{equation}
$R_{\rm max}$ is the radius of the disk. It is an almost constant density for $R<<R_{\rm max}$ and
it declines fast to zero for $R\lesssim R_{\rm max}$
In Fig. \ref{Fig:MONDexact7}, we show the rotation curve for a total mass
of $10^{11}$ $M_\odot $ and maximum radius of 20 kpc, for Newtonian gravity ($\mathbf{g}_{\rm N}$), MOND-rule ($\mathbf{g}_{\rm M-R}$) and MOND-exact ($\mathbf{g}_{\rm M-E}$) respectively, only with two iterations of the Eqs. (\ref{acc_newton}) and (\ref{rhoM-E})).

\begin{figure}
\vspace{0cm}
\centering
\includegraphics[width=9cm]{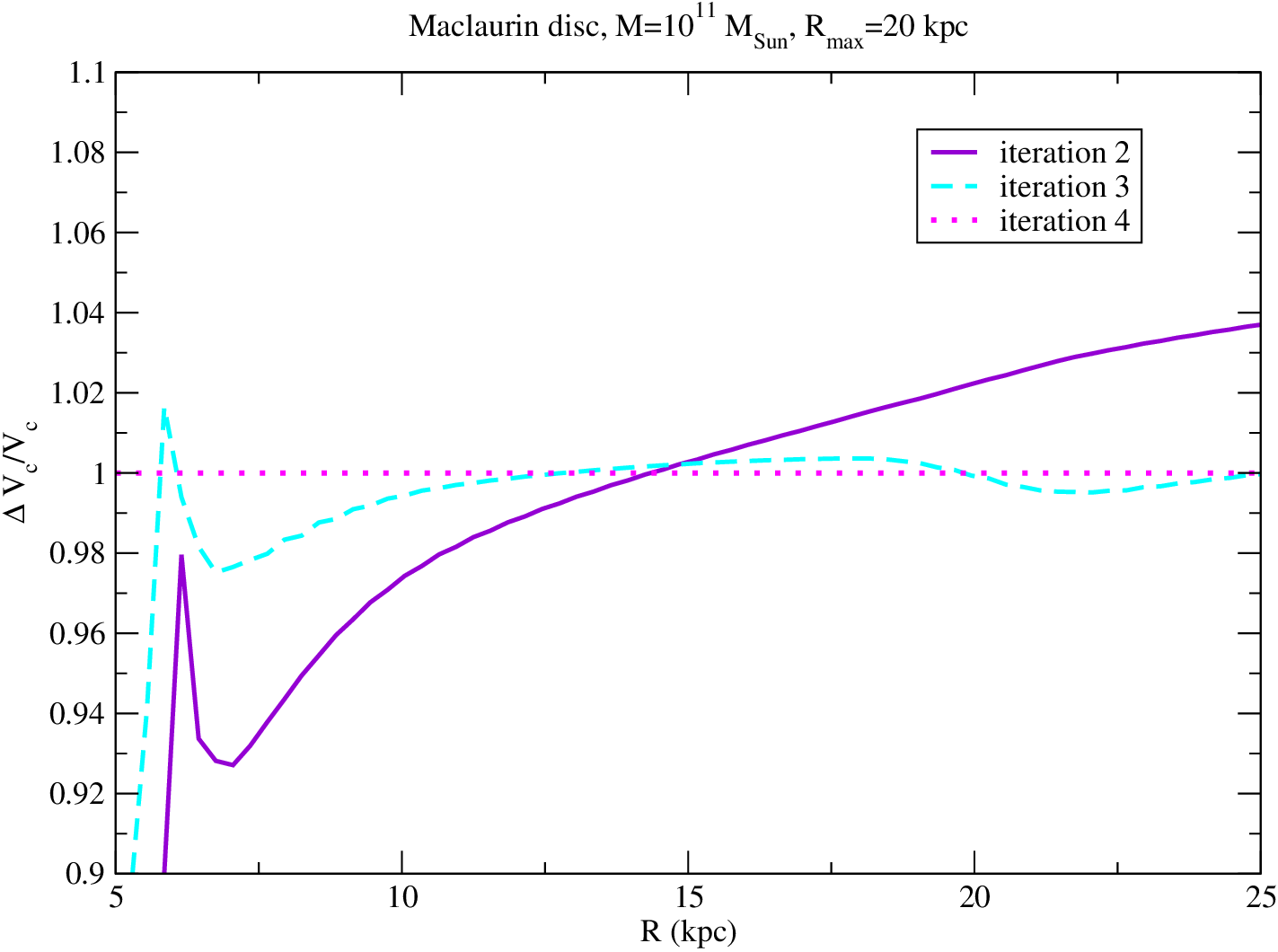}
\caption{Relative difference of the iterations 2 and 3 with respect to iteration 4 in 
the calculation of the rotation curve ($V_c=\sqrt{|g_{\rm R}|\,R}$) 
for a Maclaurin disk of $M=10^{11}$ $M_\odot $, $R_{\rm max}=20$ kpc in MOND-exact.}
\label{Fig:MONDexact8}
\end{figure}

This is a case with strong differences between $\mathbf{g}_{\rm M-R}$ and $\mathbf{g}_{\rm M-E}$, so we explore higher number iterations, to see that within iteration 3 or 4 it converges at 
$R\gtrsim 5$ kpc: see Fig. \ref{Fig:MONDexact8}. We see that the corrections of a higher iteration than
2 are of second order even in this case of differences between $\mathbf{g}_{\rm M-R}$ and $\mathbf{g}_{\rm M-E}$, which we will not take into account from now on.

\subsection{Cases with highest differences}

We have seen in the previous subsections that the MOND-rule may be very inexact with respect to MOND-exact in cases of thin disk where most of the mass is concentrated in the outer parts. The most extreme
case among those we have tested is the Maclaurin disk.

In all of the previous cases, we have set a total mass disk of $M=10^{11}$ $M_\odot$. Let us define
$R_*$ as the radius where the MOND regime is significant at more than 10\% in rotation speeds: $\left[\frac{|V_{\rm c,MOND-rule}(R_*)-V_{\rm c,Newton}(R_*)|}{V_{\rm c,Newton}(R_*)}\right]=0.1$. We also define
$r(R_*)\equiv \frac{M(R>R_*)}{M}$ (the ratio of mass at distances larger than $R_*$, where MOND is significant) and 
 $d(R_*)$ as the maximum $\forall R>R_*$ of the ratio 
$\left[\frac{|V_{\rm c,MOND-exact}(R)-V_{\rm c,MOND-rule}(R)|}{V_{\rm c,MOND-rule}(R)}\right]$. 
In previous Figs. \ref{Fig:MONDexact1}--\ref{Fig:MONDexact7}, the values of 
$R_*$ are 9.8, 10.6, 0, 7.6, and 0 kpc and
$r(R_*)$ for the five models used  are 0, 0.139, 1, 0.477, and 1, respectively, whereas the
their respective values of $d(R_*)$ are 0, 0.0390, 0.0592, 0.0474, and $>$0.70. Roughly, we
can see that the highest values of $d(R_*)$ are obtained for the values with highest mass
in the external parts within the MOND regime.
The error with respect to the algebraic solution is larger than 5\% when more than half of
the mass is in the MONDian regime, reaching in some cases $>70$\% discrepance.

\subsection{Anomalies in rotation curves}

Most spiral galaxies present an approximately flat rotation curve in the outer part, although there may be significant deviations from that behavior.  
A rising rotation curve is observed in the Andromeda galaxy, which was claimed to be 
challenging for a model
with standard dark matter models or perturbations of the galactic disk
by satellites \citep{Rui10}. However, we see here that MOND with a Mestel or Maclaurin disk naturally gives
this increase, and Newtonian for the Maclaurin disk alone; adding a dark matter halo with a strong distribution
of mass in the outer parts would reinforce this trend.

A decrease of rotation curve instead of flat curve \citep[e.g.,][]{Eil19,Zob20} in the very outer disk might be more surprising for MOND, since one expects an asymptotic limit $\lim _{r\to \infty} \mathbf{g_{M-R}}=\frac{\sqrt{G\,M\,a_0}}{r}$
and $V_c=\sqrt{|\mathbf{g_{M-R}}|r}$. In our analyses of the different disk models, we see for the MOND-exact solution that the slope is always more negative than in the MOND-rule approximation.
For instance, at $R=50$ kpc we get that $\left(\frac{dV_c}{dR}\right)_{\rm MOND-Exact}$  -0.197, -0.397, -0.212, and -0.400 km s$^{-1}$ kpc$^{-1}$ for
Figs. \ref{Fig:MONDexact2}--\ref{Fig:MONDexact7}, respectively, 
whereas $\left(\frac{dV_c}{dR}\right)_{\rm MOND-Rule}$ values are -0.087, -0.254, -0.111, and -0.165 km s$^{-1}$ kpc$^{-1}$. That is,
the negative slope at $R=50$ kpc is multiplied in the exact solution by a factor of 1.6--2.4 with respect to the approximate rule calculation. Therefore, this factor is important and should be taken into account
in a discussion about MOND plausibility in some decreasing rotation speeds.

\begin{figure}
\vspace{0cm}
\centering
\includegraphics[width=9cm]{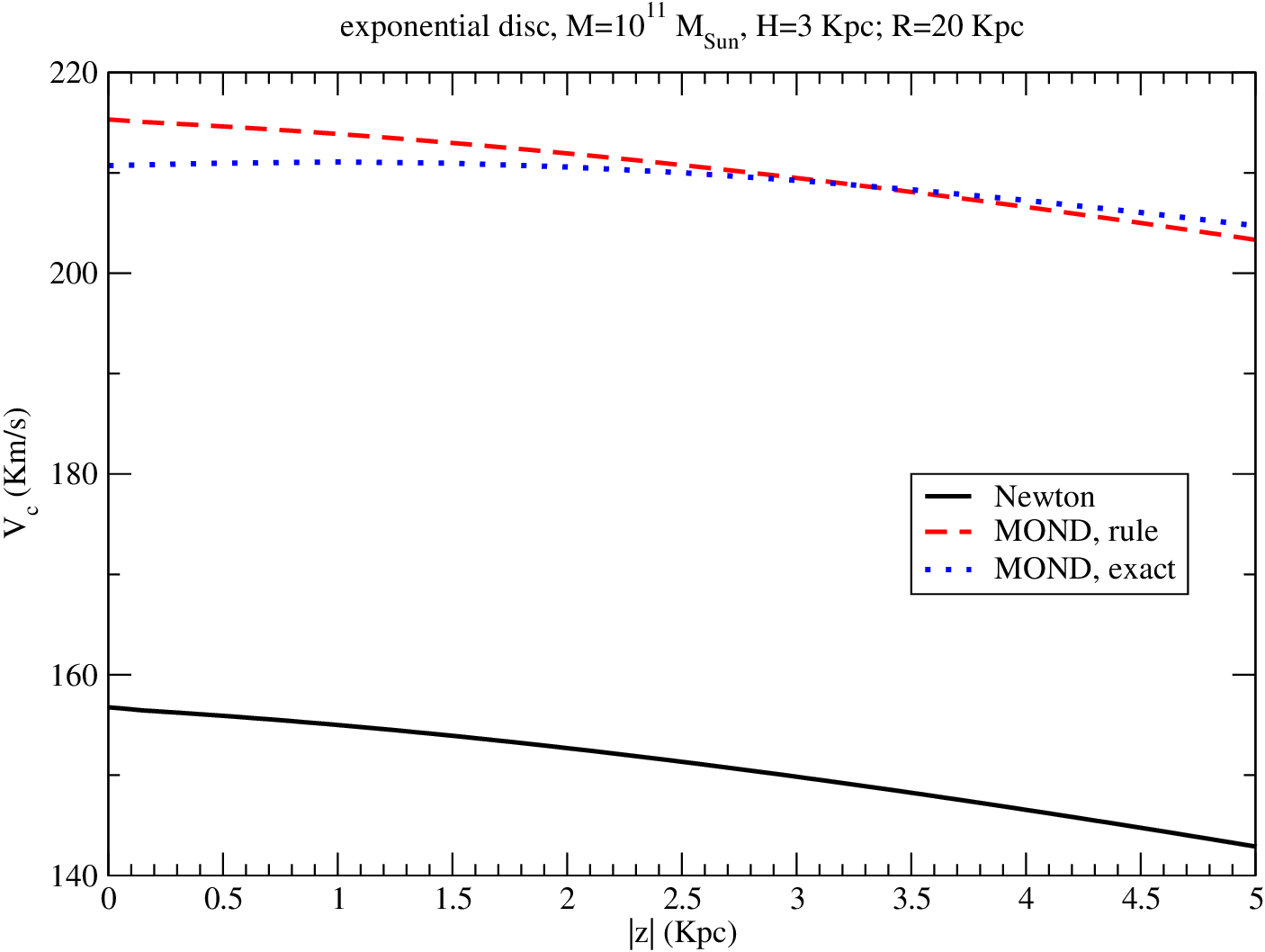}
\caption{Amplitude of the rotation curve at $R=20$ kpc 
for an exponential disk  of $M=10^{11}$ $M_\odot $, $H=3$ kpc, calculated with Newtonian gravity ($\mathbf{g}_{\rm N}$), MOND-rule ($\mathbf{g}_{\rm M-R}$), and MOND-exact ($\mathbf{g}_{\rm M-E}$), respectively.}
\label{Fig:MONDexact_z1}
\end{figure}

\begin{figure}
\vspace{0cm}
\centering
\includegraphics[width=9cm]{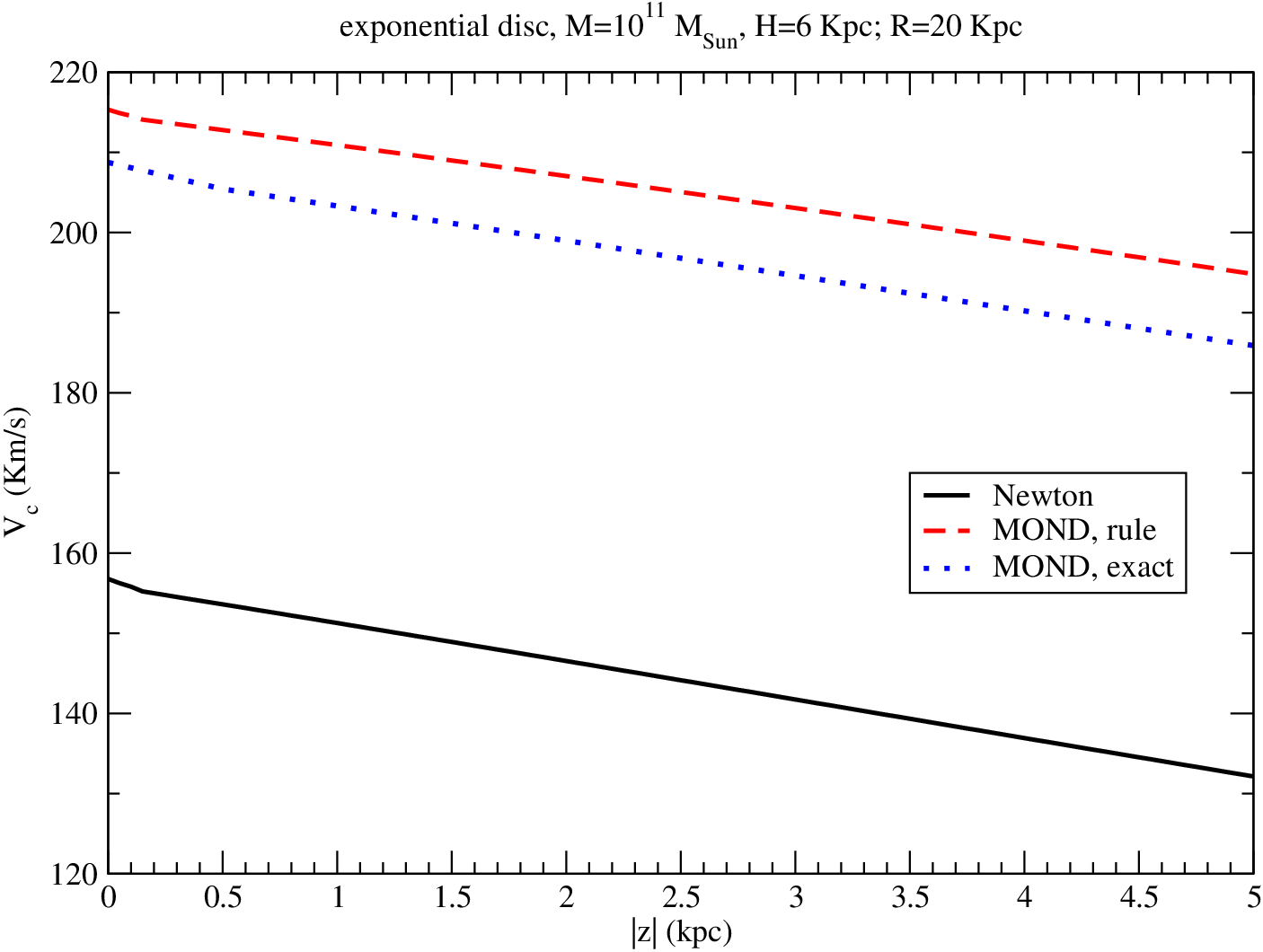}
\caption{Amplitude of the rotation curve at $R=20$ kpc 
for an exponential disk  of $M=10^{11}$ $M_\odot $, $H=6$ kpc, calculated with Newtonian gravity ($\mathbf{g}_{\rm N}$), MOND-rule ($\mathbf{g}_{\rm M-R}$), and MOND-exact ($\mathbf{g}_{\rm M-E}$), respectively.}
\label{Fig:MONDexact_z2}
\end{figure}

Another caveat in the fit of rotation curves in MOND stems from the dependence of the amplitude of the
rotation curve on the distance from the plane ($z$). In the Milky Way,  \citet{Jal10} observe that this dependence is strong and favours a Galaxy without a dark matter halo; however,
the Jeans equation to convert azimuthal velocities into rotation speed was not fully considered. A more recent analysis by \citet{Chr20} with Gaia data, taking into account the dispersion of velocities with the Jeans equation, gives a mild or negligible dependence with $z$, implying that a spherical 
component dominates. 
Let us calculate here, in our exponential disk examples, the dependence on $z$ of the rotation speed at $R=20$ kpc; the results are in Figs. \ref{Fig:MONDexact_z1} and \ref{Fig:MONDexact_z2}. We see that, for the lowest value of $H$, the MOND-exact solution gives a flat dependence of $|z|$, whereas the MOND-rule approximation
predicts a slight fall-off. In a sense, the phantom mass of MOND behaves more like a spherical distribution similar
to the halo, and this is something to be considered in the evaluation of the suitability of MOND to
fit rotation curves away from the plane.

\section{Phantom mass}
\label{.phanton}

Equation (\ref{rhoM-E}) of the exact MOND calculation has two terms of the total density
$\rho^*$: the first term is the real density, modified by a factor $\mu ^{-1}$, and the second
term is a fictitious density created by the acceleration field \citep{Mil86}.

\subsection{Central point-like mass}

In the example of a central point-like mass, as introduced in \S \ref{.central},
Eq. (\ref{rhoM-E}) gives a total density $\rho^*(r)$ with spherical symmetry (only
dependent on the distance to the center $r$) equal to

\begin{equation}
\label{roastpl}
\rho ^*(r)=\rho (r)+\frac{1}{4\pi\,G\,a_0\,\mu(x)}
\frac{d\mu }{dx}(x)
\left|g_{\rm M-E}(r)\frac{d g_{\rm M-E}(r)}{dr}\right|
,\end{equation}
\[x=\frac{|g_{\rm M-E}|(r)}{a_0}
,\]\[
g_{\rm M-E}(r)=g_{\rm M-R}(r)=\frac{G\,M}{r^2\mu(x)}
.\]
Developing this equation, and introducing Eq. (\ref{mu}) into it, we get
\begin{equation}
\label{rhoastpl}
\rho ^*(r)=\rho (r)+\frac{M}{2\pi r^3}\frac{\sqrt{1+x^{-2}}}{2+x^2}
,\end{equation}
with $\rho =0$ if $r\ne 0$. In the limit of infinite distance
\begin{equation}
\lim _{r\to \infty}\rho ^*(r)=\lim _{x\to 0}\rho ^*(r)=\frac{1}{2\pi }
\sqrt{\frac{M\,a_0}{G}}\frac{1}{r^2}
,\end{equation}
that is, it falls down inversely proportional to $r^2$. 
This is close to the decrease as $r^{-1.76\pm 0.12}$ obtained as the best halo
fit within Newtonian gravity 
derived from rotation curves \citep{Bor91}, which was interpreted as two-point 
correlation function for the galaxy background in the range of 3--350 kpc, highlighting
the coincidence of the exponent (1.76) with that of the two-point
correlation function among galaxies (1.77) \citep{Bor91}.

The total equivalent mass 
associated with this density $\rho ^*$ within a sphere of radius $r$ is
\[
M^*(r)=4\pi \int _0^r dr'\,r'^2\rho ^*(r')=M+\frac{M}{a_0}\int _\infty^{g(r)}dg'\frac{1}{\mu (x')^2}
\frac{d\mu }{dx}(x')
\]\begin{equation}
=M\left(1+\int _\infty ^{\mu (r)}d\mu'\,\frac{1}{\mu '^2}\right)=M\mu (r)^{-1}
\label{massastpl}
.\end{equation}
The total mass diverges as $r\to \infty$ and $\mu \to 0$, 
\begin{equation}
\label{Mfictinf}
\lim_{r\to \infty}M^*(r)=\sqrt{\frac{M\,a_0}{G}}r
,\end{equation} 
but only $M$ is a real mass, while
$M^*(r)-M$ stems from a fictitious density field that gravitationally behaves as Newtonian mass
but it does not correspond to any real mass. 
It is a phantom mass. 
One needs to imagine that each body generates a potential around it, which defines its gravitating mass (the phantom mass), and this potential is changed depending on which other gravitating bodies are around it and where they are, so the equivalence between gravitating mass and inertial mass is broken \citep{Wu15}. 
This concept of phantom mass as a tool to compute the MOND potential has indeed been raised to an exact concept in QUMOND, precisely motivated by earlier considerations similar to those given here but in the solar system context \citep{Mil09}. The computing of the phantom mass is the exact way to solve the QUMOND Poisson equation.

%\subsection{Implications of phantom for the large-scale}

In practice, this phantom mass does not reach infinity values, because the galaxies (to be considered point-like objects at large distances) cancel their gravitational fields in regions where other galaxies have a predominant effect. Eq. (\ref{roastpl}) is applicable in the volume where the acceleration of the galaxy is predominant, roughly on average up to a distance of half of the average separation among galaxies.
This phantom mass would be the substitute of the nonbaryonic dark mass in the standard cosmological model,
although it is not clear whether a MOND cosmology can be built, for which there are arguments in favor or
against \citep{Fel84,San98}.

%With a baryonic density of $\Omega _b=0.022h^{-2}$ \citep{Pla18}, assuming $H_0=70$ km s$^{-1}$ Mpc$^{-1}$, the 
%average density of baryonic matter is $6.1\times 10^9$ M$_\odot $ Mpc$^{-3}$. In a simple rough model
%assuming that all the mass is concentrated in aggregates of baryonic masses of $M$, we would have $\frac{6.1
%\times 10^9  M_\odot }{M}$ aggregates Mpc$^{-3}$, that is, an average separation of 
%$\left(\frac{3\,M}{4\,\pi \,6.1\times 10^9\ {\rm M}_\odot }\right)^{1/3}$ Mpc
%between these aggregates neglecting the correlations, which, following Eq. (\ref{Mfictinf}) assuming a maximum distance of half of the average separation, would have a total fictitious mass equal to
%\begin{equation}
%M^*\sim \frac{1\ {\rm Mpc}}{2}\sqrt{\frac{M\,a_0}{G}}\left(\frac{3\,M}{4\,\pi \,6.1\times 10^9\ {\rm M}_\odot }\right)^{1/3}=M\frac{5\times 10^3}{M(M\odot )^{1/6}}
%.\end{equation}
%Now, if we think that gravitational MOND effect will replace all forms of dark matter and dark energy, and only
%baryonic matter exists, while in order to preserve a flat Universe we keep a total $\Omega =1$, then
%\begin{equation}
%\frac{5\times 10^3}{M(M\odot )^{1/6}}\sim \frac{1}{\Omega _b}
%.\end{equation}
%Hence, with the above numbers $M\sim 1.3 \times 10^{14}$ M$_\odot $ and $M^*\sim 3\times 10^{15}$ M$_\odot $. This is the order of magnitude of the gravitational mass ($M^*$) of typical rich
%clusters of galaxies, which might be considered as the units of these aggregates.

\subsection{Absolute acceleration}

We could say that the idea of dependence on relative accelerations would
be impossible to sustain based on theoretical grounds. There is a historical
discussion already from the times of Helmholtz: a field cannot have a huge $N$
degrees of freedom as it would be required if the force were to depend on all of
the relative accelerations with each of the $N$ particles of the gravitational interaction. A dependence on the absolute acceleration makes more sense. Nonetheless, the concept of absolute acceleration is also problematic, since
this requires an absolute frame of reference in the universe. Which is this
absolute reference system? The cosmic microwave background radiation? 
But then the center of each galaxy has some acceleration with respect to that system. 
On the other hand, if we put the absolute
reference frame in the center of a given galaxy, how can we understand the
motions in other galaxies, which would have non-MONDian accelerations with respect to the first one? 
It is not clear either whether the acceleration is in the comoving or physical cosmological frame, and
as mentioned, it is not even clear whether we may have a MOND cosmology \citep{Fel84,San98}. For considering
the comoving frame, we would need a well-understood metric, equivalent to the one derived from general relativity. If we considered a Newtonian-like approach and paid attention only to the physical accelerations with an expansion of the universe equivalent to the standard model. we would have a relative redshift drift among galaxies with $\dot{z}\approx H_0\,z$ at low $z$ \citep{Bol19}, so the relative acceleration would be $g=c\dot{z}\approx 5.7 a_0z$, which is not negligible and can be considered in the Newtonian regime when the distance between galaxies is high ($z\gtrsim 0.17$). Something remains unclear with respect to the concept of absolute acceleration. Certainly, the concepts of MOND are slippery, but we may
forget about the conceptual theoretical problems and see whether the phenomenological rules can be applied
at least within one galaxy. 

\subsection{MOND and External Field Effect}

There are some attempts to clarify the question of the superposition of
fields, distinguishing the external field effect (\citet{Mil83a}; \citet[Section 6.3]{Fam12}) and the internal field effect and claiming that the internal accelerations of the subsystem are irrelevant to how that subsystem responds to an external field. Only the field of the parent system at the position of the center of mass of the subsystem is relevant to that. Only the center of mass of such systems matters to determining their orbits in MOND, not their internal structure, nor the magnitude of their internal accelerations.
%``the internal accelerations of the sub-system are irrelevant to how that sub-system responds to an external field. Only the field of the parent system at the position of the center of mass of the sub-sytem is relevant to that. (...) Only the center of mass of such systems matters to determining their orbits in MOND, not their internal structure, nor the magnitude of their internal accelerations''.
%\footnote{http://astroweb.case.edu/ssm/mond/EFE.html by S. S. McGaugh.}. 
MOND has to be described by a nonlinear theory. This basically means that the acceleration endowed by two bodies to a third is not the (vectorial) sum of the individual accelerations produced by each separately. 
%There is not an answer on how come the combined, center-of-mass acceleration.
%``MOND has to be described by a nonlinear theory. This basically means that the acceleration endowed by two bodies to a third is not the (vectorial) sum of the individual accelerations produced by each separately. (...) how come the combined, center-of-mass acceleration in the field of B is MONDian? I don't actually have a non-technical, intuitive, or heuristic answer. This, in fact, happens differently in different MOND theories.''.
%\footnote{http://astroweb.case.edu/ssm/mond/milgromonefe.html by S. S. McGaugh, mentioning explanations given by M. Milgrom.} 
%We may understand the non-linearity, but this distinction between internal and external acceleration does not look clear.

%MOND has indeed had serious attempts of mathematical formalization of these 
%problems \citep{Bek84,Fam12}, 
%The classical papers that constituted the seed of MOND idea \citep{Mil83a,Mil83b,Mil83c} already showed on phenomenological grounds 
%that the internal dynamics of the systems should be approximately Newtonian
%even if the relative internal accelerations are much smaller than $a_0$.
\citet{Bek84} derived mathematically two consequences of the superposition
of fields: 1) that the acceleration of the center of mass in a system much smaller
than the source of the external field (e.g., a star subject to the external
field produced by the rest of the galaxy) follows the acceleration imposed by
the external object when the radius of the system trends to infinity \citep[Section IV]{Bek84}; and
2) if the external field is Newtonian, assuming also a large enough radius, the internal field of an
arbitrary mass in the system is Newtonian even though the internal accelerations
are much smaller than $a_0$ \citep[Sect. V]{Bek84}.
%The violation of the strong principle of equivalence in MOND makes that the internal dynamics of a system
%in a constant external field is different from that of the same system in the
%absence of it \citep{Bek84}, 
Nonetheless, it is not clear what can be considered external
or internal. Given an atom that feels two gravitational accelerations, how can it distinguish whether the force comes from a nearby source or a distant source? Or is MOND also dependent on distance apart from the dependence on the acceleration? In principle, it is not.
Certainly, a particle only feels a total acceleration without distinguishing where it comes from. The distinction between internal and external field is not something a particle is aware of.
This was indeed the apparent paradox that one of us  introduced \citep{Lop18}. However, these suspicions of
contradiction were not correct.

From our analysis, the explanations of these cases are as follows:
%\begin{enumerate}
%\item 
On the Earth, for instance, the phantom mass [the second term in Eq. (\ref{rhoM-E})] of the fields created
by the atom or the small particle is canceled, due to the action of the strong field of the Earth in which it
is embedded. Also, for a multiple star system in the inner disk of a galaxy within the Newtonian regime, similar argumentations can be given. Therefore, we cannot explore MOND effects in binary stars or globular clusters in the inner Galaxy, as apparently found in some observations \citep[e.g.,][]{Sca17}, unless this MOND logic does not apply for some other reason.
%\item 

However, on the interior of a star in the outer disk of a galaxy, the self-gravity cannot cancel the MONDian phantom mass created by the center of the galaxy because almost all of the space filled by this mass is far from the volume of the Newtonian regime owing to self-gravity of the star, which is negligible in comparison with the total volume. Therefore, MONDian acceleration is applied over all of the atoms of the star, and consequently the center of mass of the star follows a MONDian dynamics.
%\item 
On a binary or multiple-star system in the outer disk of a galaxy, similar argumentation can be given.

\section{Conclusions}
\label{.conclus}

While the AQUAL \citep{Bek84} or QMOND \citep{Mil10} theoretical frames for the definition of the MOND field were developed and applied through numerical simulations with Poisson solver codes (usually $N$-body and hydrodynamics codes are used) to galactic dynamics problems \citep{Bra99,Tir07,Lon09,Ang12,Can15,Lug15,Ban18b}, here we have followed the proposal by \citet{Mil86} to develop
a semianalytical algorithm based on the iteration of some  
analytical expressions that allow the calculation of the MOND accelerations with any mass distribution. As realized by \citet{Mil86}, we see that the application of this formalism is equivalent to the creation of a fictitious phantom mass whose field may be used in a Newtonian way to calculate the MOND accelerations.
This analytical recipe was already used, for instance, by \citet{Mil09} in the solar system context, an approach that has actually led \citet{Mil10} to propose the exact QUMOND theory back.
QUMOND \citep{Mil10} might also be used, although here we have focused on AQUAL solutions. In any case, the conclusion with QUMOND would be similar, since, despite using different field equations, QUMOND and AQUAL give rather similar results, as demonstrated both numerically \citep{Can15,Can16} and analytically \citep{Ban18a}.

%We also diskuss some unclear items related to MOND and how they can be clarified with the phantom mass in the exact solution. We state in a clearer way than in previous works that the external field effect is applicable
%at short distances, so we can understand why the Newtonian field of the Earth does not allow to apply MOND to nearby sources, whereas the Newtonian field of the components of a star or a multiple system does not cancel the MOND dynamics when the origin of the field stems from a distant source.

An interesting application of the approach presented here for solving  
the MOND equation is the treatment of the problem of a system orbiting  
in the MOND region of a larger system. If the smaller system is all  
within the internal Newtonian region, it is clear that the internal  
dynamic is Newtonian while that of its center of mass is Mondian.  
However, when the internal dynamics is itself in the MONDian regime,  
the internal and external contributions do not separate in this  
simple manner, particularly when the internal dynamics is just  
entering the MONDian regime. With
the equation used in this work, however, this problem can be  
accurately used, taking for the phantom mass that given by MOND simple  
rule as a first guess.

We extend the analyses of \citet{Bra95} to show how much is the difference between the ``exact MOND'' calculation using AQUAL \citep{Bek84} and the usual algebraic rule of calculating the MOND force as equal to the Newtonian one divided by some factor $\mu $ when applied to the calculation of rotation curves, as usually done. We corroborate that in most cases the differences are small, $\lesssim 10$\%. However, some density distributions with large fractions of mass in the outer part, such as Maclaurin disks, may show much higher differences.
The slope of the rotation speed in the dependence with the radius can also be changed by a large factor; therefore,
any discussion about the compatibility of an outward decrease of rotation speed in the outer disk instead of an
expected asymptotical flat shape, should be analyzed in terms of these exact calculations if we want to test MOND.
Moreover, the dependence of the rotation speed amplitude on the vertical distance from the Galactic plane is 
also significantly different between the exact and the approximate rule solution.
These subtleties in the calculation of MOND with the exact solution are important when using high-precision data.

In the past years, data with large precision have allowed the determination of rotation curves with high accuracy. Within this context of precision dynamics, it is necessary to apply exact calculations for the rotation curves, rather than simple rule approximations of acceleration as the Newtonian one divided by some factor $\mu $. Therefore, we encourage astronomers to use the exact rule given in this paper, especially when
one tries to examine small subtleties that might allow the rejection of MOND or dark matter hypotheses.

A three-body problem with two very massive body systems (instead of one very massive source 
that we have considered here for the external field) plus a
small mass body would be an interesting exercise to carry out, which was not analyzed here since it is beyond
the scope of this paper. One might, for instance, consider how is the combined phantom mass of the Milky Way and Andromeda galaxy together, in order to calculate the exact MOND dynamics over a minor object like a dwarf galaxy. 
%The application of the concept of phantom mass in the large-scale structure using simple exact solutions of AQUAL like we did here is a field that warranties further
%research. At present, with simple order of magnitudes calculations, we derive that formation aggregates of gravitational masses of $\sim 3\times 10^{15}$ M$_\odot $ (including real and phantom mass), typical of rich cluster of galaxies, are enough to make $\Omega =1$ in a MOND Universe without dark matter and dark energy.

\begin{acknowledgements}
Thanks are given to Stacy McGaugh, Indranil Banik, Pavel Kroupa, and Riccardo Scarpa 
for comments and suggestions on a draft of this work. Thanks are given to the anonymous referee for helpful comments.
M.L.-C. was supported by the grant PGC-2018-102249-B-100 of
the Spanish Ministry of Economy and Competitiveness (MINECO).
\end{acknowledgements}

\end{document}